\documentclass[amssymb,amsmath,aps,showpacs,twocolumn]{revtex4}

\newcommand{\lsim}{\lesssim}
\newcommand{\gsim}{\gtrsim}
\newcommand{\hin}{h_{inv}}
\newcommand{\pt}{p_T\!\!\!\!\!\! /\,\,}

\def\lsim{\mathrel{\raise.3ex\hbox{$<$\kern-.75em\lower1ex\hbox{$\sim$}}}}
\def\gsim{\mathrel{\raise.3ex\hbox{$>$\kern-.75em\lower1ex\hbox{$\sim$}}}}

\def\beq{\begin{equation}}
\def\eeq{\end{equation}}
\def\beqn{\begin{eqnarray}}
\def\eeqn{\end{eqnarray}}
\def\bea{\begin{eqnarray}}
\def\eea{\end{eqnarray}}
\def\be{\begin{equation}}
\def\ee{\end{equation}}
\newcommand{\fslash}[1]{{#1 \kern -0.7em/ \kern 0.1em}}
\def\ptmiss{\fslash{P}_T}

\begin{document}

\voffset 1.25cm

\title{ Detecting an invisible Higgs boson at Fermilab Tevatron and CERN LHC }
\author{ Shou-hua Zhu}
\affiliation{Institute of Theoretical Physics, School of Physics,
Peking University, Beijing 100871, China}

\date{\today}

\begin{abstract}

In this paper, we study the observability of an invisible Higgs
boson at Fermilab Tevatron and CERN LHC through the production
channel $ q \bar{q} \rightarrow Z H \rightarrow \ell^+\ell^-+
\ptmiss $, where $\ptmiss$ is reconstructed from the $\ell^+\ell^-$
with $\ell=e$ or $\mu$. A new strategy is proposed to eliminate the
largest irreducible background, namely $ q \bar{q} \rightarrow
Z(\rightarrow \ell^+\ell^-) Z(\rightarrow \nu \bar\nu)$. This
strategy utilizes the precise measurements of $ q \bar{q}
\rightarrow Z(\rightarrow \ell^+\ell^-) Z(\rightarrow
\ell^+\ell^-)$. For $m_H=120$ GeV and with luminosity $30 fb^{-1}$
at Tevatron, a $5\sigma$ observation of the invisible Higgs boson is
possible. For $m_H=114 \sim 140$ GeV with only $10 fb^{-1}$
luminosity at LHC, a discovery signal over $5\sigma$ can be
achieved.

\end{abstract}

\maketitle

\section{Introduction}

Understanding the mechanism of electroweak symmetry breaking (EWSB)
is a primary goal of the Fermilab Tevatron, CERN LHC and the
proposed ILC. In the standard model (SM) of high energy physics,
EWSB is realized via a weak-doublet fundamental Higgs field. After
EWSB spontaneously, namely Higgs field acquiring  a vacuum
expectation value (VEV), only one neutral Higgs boson is left in
particle spectrum. The Higgs boson mass is theoretical unknown
within the SM. Therefore searching all mass region is necessary and
great efforts have been put on it since the establishment of the SM.
The latest direct search at LEP sets the lower bound of SM Higgs
boson of $114.4$ GeV at 95\% confidence level (CL)
\cite{Barate:2003sz}. The Higgs boson can also affect electro-weak
observables through radiative corrections. Therefore precise
measurements of these observables can predict the Higgs boson mass.
Studies show that data from LEP, SLD and Tevatron are in good
agreement with SM predictions \footnote{The notorious three
3-$\sigma$ anomalies \cite{LEP-latest-talk} may indicate new
dynamics beyond the SM, moreover even without these anomalies the
global fit shows certain tension with direct search limit at LEP
\cite{Chanowitz:2001bv}.}.  Based on the global fit of those data,
the Higgs boson mass is predicted to be $m_H=98^{+52}_{-36}$ GeV and
$m_H <208$ GeV at 95\% CL using latest preliminary top quark mass
$m_t=174.3\pm 3.4$ GeV \cite{Juste:2005dg}.


The Higgs boson decay width in the SM varies dramatically within the
experiments preferable mass region $114.4\sim 208$ GeV. For example,
for $m_H=120$ GeV, $\Gamma(H)\simeq 3.65$ MeV while for $m_H=200$
GeV, $\Gamma(H)\simeq 1.425$ GeV \cite{Djouadi:2005gi}. The tiny
decay width for light Higgs ($< 2 m_W$) is due to the suppressed
coupling among Higgs boson and fermion which is proportional to
$m_f/m_W$ with $m_f$ the light fermion mass.
Therefore the {\em light} Higgs boson ($< 2 m_W$) can possibly decay
into non-SM particles  with large branching ratio. For example the
Higgs boson may decay dominantly into scalar dark matter in the
simplest cold dark matter model \cite{myself}. Study of Ref.
\cite{myself} shows that the correct cold dark matter relic
abundance within $3\sigma$ uncertainty ($ 0.093 < \Omega_{dm} h^2 <
1.129 $) and experimentally allowed Higgs boson mass ($114.4 \le m_H
\le 208$ GeV) constrain the scalar dark matter mass within $48 \le
m_S \le 78$ GeV. This result is in excellent agreement with that of
W.~de Boer et.al. ($50 \sim 100$ GeV) \cite{deBoer:2005tm}. Such
kind of dark matter annihilation can account for the observed gamma
rays excess ($10\sigma$) at EGRET for energies above 1 GeV in
comparison with the expectations from conventional Galactic models.
The most important phenomenological consequence of this model is
that the Higgs boson decays dominantly into scalar cold dark matter
if its mass lies within $48 \sim 64$ GeV. In Ref.
\cite{Boehm:2003bt}
 $O(1\sim 100)$ MeV scalar dark matter was proposed to
account for the observation of a 511 KeV bright $\gamma$-ray line
from the galactic bulge \cite{511KeV}. From theoretical point of
view, containing a stable singlet scalar which interacts possibly
with SM Higgs boson, is a generic feature of  models of scalar dark
matter \cite{myself,SterileScalar}.
In other models beyond the SM, it is not rare that the Higgs boson
can decay into dark matter. For example in supersymmetical models
with R-parity, the Higgs boson can decay into pair of neutralino if
kinematically allowed.

Another reason why light Higgs boson is especially interesting is
due to the aspect of experiments. It is known for a long time that
hadronic asymmetry measurements, namely $A_{FB}^b, A_{FB}^c$ and
$Q_{FB}$ prefer the heavier Higgs mass with central value around 400
GeV, while the leptonic ones, $m_W$,$F_Z$ and $R_\ell$ prefer the
very light Higgs which already indicates certain tension with direct
search limit 114.4 GeV \cite{Chanowitz:2001bv}. While such situation
maybe  totally due to statistical fluctuations, it is not
unreasonable to suspect that some unknown systematic errors lie in
hadronic asymmetry measurements. If this is true, the SM Higgs boson
tends to lie just above the current direct search limit, otherwise
the tension will become stronger.

In this paper we will study invisible Higgs boson in $ZH$ associated
production channel in which Higgs boson decays {\em invisibly}, i.e.
we can't tag Higgs decay products, and $Z \rightarrow \ell^+\ell^-$
with $\ell=e,\mu$. The signal is $\ell^+ \ell^- \ptmiss$ where
$\ptmiss$ is reconstructed from the $\ell^+\ell^-$. This channel has
been widely discussed in literature \cite{Choudhury:1993hv,
Frederiksen:1994me,Martin:1999qf,Godbole:2003it,Davoudiasl:2004aj}.
As one of the characteristics of this mode, we can't get Higgs boson
invariant mass peak out of continuum background. Therefore it is
quite interesting to know how to get Higgs boson mass information
from the experimental measurements. As pointed out by Ref.
\cite{Davoudiasl:2004aj}, this process may provide an interesting
handle on the Higgs boson mass at LHC. The Higgs mass can be
extracted from the production cross section and the uncertainty is
35-50 GeV (15-20 GeV) with integrated luminosity $10 (100) fb^{-1}$
at LHC \cite{Davoudiasl:2004aj} \footnote{The precise measurement of
$ZZ \rightarrow 4\ell$ can reduce the largest ZZ background, and it
may also improve the Higgs mass determination.}. Therefore, precise
understanding of backgrounds is essential. At hadron colliders the
precise predictions for backgrounds are commonly thought difficult
because of uncertainty in parton distribution function (PDF), large
QCD radiative corrections etc. However for this channel, we only
care about charged leptons final states, and the backgrounds seem to
be less severe than those of hadronic final states. Another
interesting question is how to suppress background efficiently. In
literature, lots of techniques are proposed \cite{Choudhury:1993hv,
Frederiksen:1994me,Martin:1999qf,Godbole:2003it,Davoudiasl:2004aj}.
However the largest irreducible ZZ background, with one Z decays
into neutrinos and the other charged lepton pair, can't be
eliminated by kinematical cuts. In this paper, we will propose a
method, i.e. utilizing the precise measurement of $ Z(\rightarrow
\ell^+\ell^-) Z(\rightarrow \ell^+\ell^-)$, to reduce the largest
irreducible ZZ background.

\section{Detail simulation}

In this paper, we consider the production of Higgs boson in
association with a $Z$ boson, and Higgs decays 100\% invisibly.
Therefore the signal is:
\begin{equation}
  p \, p({\rm or\ } \bar p) \to Z(\to \ell^+\ell^-) + \hin \, \, ; \qquad \ell = e,
  \mu.
  \label{pptozh}
\end{equation}

As the signal is $\ell^+ \ell^- \ptmiss$ where $\ptmiss$
is reconstructed from the $\ell^+\ell^-$, the most significant sources
of background are
\begin{eqnarray}
 & Z (\to \ell^+ \ell^-) Z (\to \nu {\bar \nu}), \
  W^+ (\to \ell^+ \nu) W^- (\to \ell^- {\bar\nu}), & \nonumber\\
  &Z (\to \ell^+ \ell^-)  W (\to \ell \nu),&
  \label{bkg}
\end{eqnarray}
with the lepton  ($e,\mu$ and $\tau$, and only here we consider
$\tau$ lepton) from the $W$ decay in $ZW$ missed, and $Z + {\rm
jets}$ final states with fake
$\ptmiss$~\cite{Frederiksen:1994me,Martin:1999qf}.  We simulate the
signal and the three backgrounds in Eq. (\ref{bkg}) using Pythia
\cite{Sjostrand:2003wg} without initial and final QCD and QED
radiation corrections. In order to reduce the fake $\ptmiss$
background to an insignificant level, we set $\ptmiss > 85, 100$ GeV
for LHC \cite{Davoudiasl:2004aj} and $\ptmiss > 55, 75, 100$ GeV for
Tevatron \cite{Martin:1999qf}.

We adopt the following ``LHC cuts'' from \cite{Davoudiasl:2004aj}:
\begin{equation}
    p_T(\ell^{\pm}) > 10 \ {\rm GeV}, \
    |\eta(\ell^\pm)| < 2.5, \
    \Delta R(\ell^+\ell^-) > 0.4,
\label{lhc1}
\end{equation}
where $\eta$ denotes pseudo-rapidity and $\Delta R$ is the
separation between the two particles in the detector, $\Delta R
\equiv \sqrt{(\Delta \eta)^2 + (\Delta \phi)^2}$; $\phi$ is the
azimuthal angle.
The $W W$ background can be largely eliminated by
\begin{equation}
    |m_{\ell^+ \ell^-}-m_Z|<10 \ {\rm GeV,} \ \ \ \
    \Delta\phi_{\ell^+ \ell^-}<2.5.
\label{lhc2}
\end{equation}
In order to reduce the third background in Eq. (\ref{bkg}), we veto
events with
\begin{equation}
p_T(\ell)> 10  \ {\rm GeV}, \ \ \ \ |\eta(\ell)|<3.0.
\label{lhc3}
\end{equation}

At Tevatron we adopt the following ``Tevatron cuts''
from \cite{Martin:1999qf}:
\begin{eqnarray}
   & p_T(\ell^{\pm}) > 12 \ {\rm GeV}, \
    |\eta(\ell^\pm)| < 2.0, \
    \Delta R(\ell^+\ell^-) > 0.4, &
\label{tev1} \\
   & |m_{\ell^+ \ell^-}-m_Z|<7 \ {\rm GeV,} \ \ \ \
    \Delta\phi_{\ell^+ \ell^-}<2.7,
\label{tev2}
\end{eqnarray}
and veto events with
\begin{equation}
p_T(\ell)> 10  \ {\rm GeV}, \ \ \ \ |\eta(\ell)|<2.5.
\label{tev3}
\end{equation}

The signal and background are shown in Fig. 1 and 2 as a function of
$\ptmiss$ for $e^+e^-$ final states at Tevatron and LHC. Our results
are consistent with those of Ref. \cite{Martin:1999qf} and
\cite{Davoudiasl:2004aj}.

\begin{figure}[thb]
\vbox{\kern2.5in\includegraphics{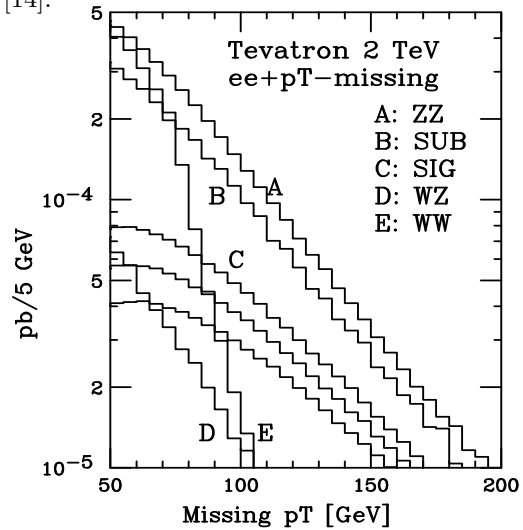}}
\caption{ Missing
$p_T$ distribution for $Z(\to e^+e^-)+\hin$ signal and backgrounds at Tevatron with
$\sqrt{s}=2$ TeV after applying the
"Tevatron cuts" in Eqs.~(\ref{tev1}), (\ref{tev2}) and (\ref{tev3}).
 Here "C" represents signal with $m_h = 120$, 130 and 140 GeV from top to bottom, and
 "SUB" stands for $R \times \sigma_{4\ell}/2$ (see text).
}
\label{tev}
\end{figure}

\begin{figure}[thb]
\vbox{\kern2.5in\includegraphics{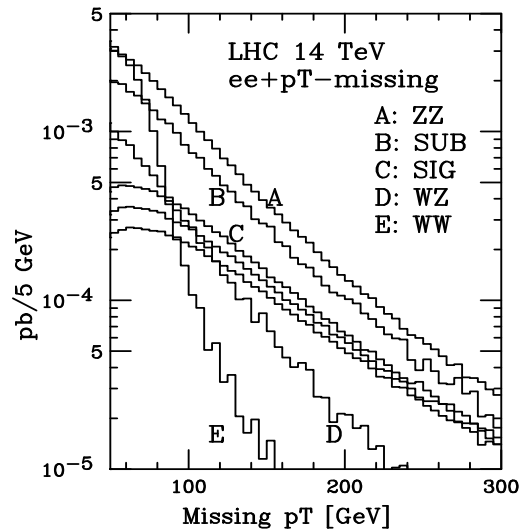}} \caption{ Same conventions with
figure \ref{tev} but at the LHC with $\sqrt{s}=14$ TeV after
applying the "LHC cuts" in Eqs.~(\ref{lhc1}), (\ref{lhc2}) and
(\ref{lhc3}).}
\end{figure}

After imposing cuts in Eqs. (\ref{lhc1}-\ref{tev3}), the remaining
largest irreducible background is ZZ, as shown in Fig. 1 and 2. This
background cannot be eliminated via kinematical cuts. Therefore we
propose to utilize the measurement of $ Z(\rightarrow \ell^+\ell^-)
Z(\rightarrow \ell^+\ell^-)$ (in short $ZZ \rightarrow 4\ell$
hereafter) to reduce this background. This idea is quite natural
because of two reasons. The first one is that
 $ZZ \rightarrow 4\ell$ is almost background
free\cite{DiBoson} \footnote{Due to the large $gg$ luminosity at
LHC, $gg \rightarrow H \rightarrow Z Z^{(*)} \rightarrow 4\ell$ can
act as the background to $ZZ\rightarrow 4\ell$ especially for
heavier Higgs, say 140 GeV. However we can apply invariant mass cut
to eliminate such kind of background.} due to the excellent leptonic
reconstruction efficiency and mass resolution of Z boson (decays to
a pair of charged leptons). For example with integrated luminosity
30 $fb^{-1}$ cross section of this channel can be measured to an
accuracy of $5\%$ at LHC \cite{DiBoson}. The second reason is that
ZZ background and $ZZ\rightarrow 4\ell$ share almost the same
kinematics, the same higher order QCD radiative correction and PDF
uncertainties, as well as the uncertainty of luminosity etc. We can
then get improved ZZ background by subtracting contributions from
$ZZ\rightarrow 4\ell$
\begin{eqnarray}
\sigma_{bkg}^{ZZ, improved}=\sigma_{bkg}^{ZZ}-R\times \sigma_{4\ell}
\label{sub}
\end{eqnarray}
where $\sigma_{4\ell}$ is the cross section for $ZZ \rightarrow
4\ell$ with each pair of leptons satisfying "LHC cuts" or "Tevatron
cuts" in Eqs. (\ref{lhc1}-\ref{tev3}). Actually $ZZ\rightarrow
4\ell$ event does not have missing $P_T$ if we run through all the
final state particles. However in our case $\ptmiss$ is
reconstructed via the momentum of charged lepton pair, invariance
mass of which is effectively around $m_Z$. Thus $ZZ\rightarrow
4\ell$ is the extra data set which is not included in that of
$\ell^+\ell^- \ptmiss$. Note that the purity of $ZZ \rightarrow
4\ell$ is the most crucial factor for the subtraction purpose. Thus
besides "LHC cuts" or "Tevatron cuts" (e.g. invariance mass of each
pair is effectively around $m_Z$), we can further require that
missing $P_T$ of four leptons should be small, in order to guarantee
that they do come from $ZZ$. The detail study of $ZZ \rightarrow
4\ell$ can be found in Ref. \cite{DiBoson}. In Eq. (\ref{sub}) $R$
is the ratio defined as
\begin{eqnarray}
R=\frac{2}{3} \frac{\sum_{i=1}^{3}Br(Z\rightarrow \nu_i \bar \nu_i)}{Br(Z \rightarrow
e^+e^-)}.
\end{eqnarray}
It is obvious that $\sigma_{bkg}^{ZZ, improved} \approx 0$ if we can
measure all 4 charged lepton final states in any kinematical region.
Though this ratio  at tree level can be expressed as, omitting the
final state lepton mass,
\begin{eqnarray}
R=2\times \frac{(g_V^{(\nu)})^2+(g_A^{(\nu)})^2}{(g_V^{(\ell)})^2
+(g_A^{(\ell)})^2}
\end{eqnarray}
with $g_V^{(f)}=T_{w3}^{(f)}-2 \sin^2\theta_W Q^{(f)}$ and
$g_A^{(f)}=T_{w3}^{(f)}$. The higher order radiative corrections can
be included by replacing the couplings $g_V$ and $g_A$ by the
effective ones. In this paper we alternatively adopt the
experimental central values as input: $\Gamma(Z\rightarrow\ {\rm
inv})= 499$ MeV and $\Gamma(Z\rightarrow \ell^+\ell^-)=83.984$ MeV
\cite{Eidelman:2004wy}, and get
\begin{eqnarray}
R=\frac{2}{3} \frac{\Gamma(Z\rightarrow\ {\rm inv})}{\Gamma(Z\rightarrow \ell^+\ell^-)}=3.961.
\end{eqnarray}

In Fig. 1 and 2, we also show the $\ptmiss$ distribution for $4\ell$
process (in fact, one can reconstruct two equal $\ptmiss$ based on
two pair of leptons, and invariant mass of each lepton pair is
around Z mass peak). The shape of ZZ background and $ZZ \rightarrow
4\ell$ process is very similar and the difference represents the
incomplete cancelation due to the kinematical cuts on charged
leptons.

Our results for the background and signal cross sections are
tabulated in Table~\ref{table1} for Tevatron and \ref{table2}
for LHC.  The corresponding signal to
background ratio, $S/B$, and significance, $S/\sqrt B$, are
tabulated in Table~\ref{table3} for Tevatron and \ref{table4}
for LHC.

\begin{table}[htb]
\begin{tabular}{l|cccc}
\hline \hline
$\pt$ cut & B($ZZ$) & improved B($ZZ$) & B($WW$) & B($ZW$)  \\
\hline
55 GeV & 6.73 & 1.78 fb & 2.96 fb & 0.70 fb  \\
75 GeV & 3.96 & 0.97 fb & 0.71 fb & 0.35 fb  \\
100 GeV & 1.96 & 0.45 fb & 0.10 fb & 0.15 fb  \\
\hline \hline
 & \multicolumn{3}{c}{S($Z+\hin$)}  \\
 $\pt$ cut &  $m_h = 120$~~~ & 130~~~  & 140 GeV  \\
\hline
 55 GeV &   2.07 fb & 1.62 fb & 1.27 fb  \\
 75 GeV &   1.46 fb & 1.18 fb & 0.95 fb  \\
 100 GeV &   0.88 fb & 0.73 fb & 0.61 fb \\
\hline \hline
\end{tabular}
\caption{ Background and signal cross sections for associated $Z(\to
\ell^+\ell^-)+\hin$ production at the Tevatron, combining the $ee$ and
$\mu\mu$ channels.
} \label{table1}
\end{table}

\begin{table}[htb]
\begin{tabular}{l|cccc}
\hline \hline
$\pt$ cut & B($ZZ$) & improved B($ZZ$) & B($WW$) & B($ZW$)  \\
\hline
85 GeV & 30.4 & 9.1 fb & 2.6 fb & 6.2 fb  \\
100 GeV & 22.0 & 6.4 fb & 1.1 fb & 4.1 fb  \\
\hline \hline
  & \multicolumn{3}{c}{S($Z+\hin$)}  \\
$\pt$ cut  &  $m_h = 120$~~~ & 130~~~  & 140 GeV  \\
\hline
 85 GeV &   10.5 fb & 8.8 fb & 7.4 fb  \\
 100 GeV &   8.3 fb & 7.1 fb & 6.0 fb \\
\hline \hline
\end{tabular}
\caption{
Same with Table~\ref{table1} but at the LHC.
} \label{table2}
\end{table}

\begin{table}[htb]
\begin{tabular}{l|ccc}
\hline \hline
         &  \multicolumn{3}{c}{$m_h = 120$ GeV} \\
$\pt$ cut & S/B & S/$\sqrt{\rm B}$ (10 fb$^{-1}$)
    & S/$\sqrt{\rm B}$ (30 fb$^{-1}$)\\
\hline
55 GeV & 0.38(0.25)        & 2.81(2.26)       & 4.86(3.91)       \\
75 GeV & 0.72(0.29)        & 3.20(2.06)      & 5.61(3.57)       \\
100 GeV & 1.26(0.40)       & 3.32(1.87)       & 5.76(3.24)        \\
\hline \hline
         &  \multicolumn{3}{c}{$m_h = 130$ GeV} \\
$\pt$ cut & S/B & S/$\sqrt{\rm B}$ (10 fb$^{-1}$)
    & S/$\sqrt{\rm B}$ (30 fb$^{-1}$)\\
\hline
55 GeV & 0.30(0.19)        & 2.20(1.77)       & 3.80(3.06)       \\
75 GeV & 0.58(0.24)        & 2.62(1.67)       & 4.54(2.88)       \\
100 GeV & 1.04(0.33)       & 2.76(1.55)       & 4.78(2.70)        \\
\hline \hline
         &  \multicolumn{3}{c}{$m_h = 140$ GeV} \\
$\pt$ cut & S/B & S/$\sqrt{\rm B}$ (10 fb$^{-1}$)
    & S/$\sqrt{\rm B}$ (30 fb$^{-1}$)\\
\hline
55 GeV & 0.23(0.15)        & 1.72(1.39)       & 2.98(2.40)       \\
75 GeV & 0.47(0.19)        & 2.11(1.34)       & 3.65(2.32)       \\
100 GeV &  0.87(0.28)     &  2.31(1.30)     &  3.99(2.25)       \\
\hline \hline
\end{tabular}
\caption{ Signal significance for associated $Z(\to
\ell^+\ell^-)+\hin$ production at the Tevatron, combining the $ee$ and
$\mu\mu$ channels. The numbers in the parentheses correspond to
ZZ background without improvement (see text).
} \label{table3}
\end{table}

\begin{table}[htb]
\begin{tabular}{l|ccc}
\hline \hline
         &  \multicolumn{3}{c}{$m_h = 120$ GeV} \\
$\pt$ cut & S/B & S/$\sqrt{\rm B}$ (10 fb$^{-1}$)
    & S/$\sqrt{\rm B}$ (30 fb$^{-1}$)\\
\hline
85 GeV & 0.59(0.27)        & 7.8(5.3)       & 13.6(9.2)       \\
100 GeV & 0.72(0.31)       & 7.7(5.0)      & 13.3(8.7)        \\
\hline \hline
         &  \multicolumn{3}{c}{$m_h = 130$ GeV} \\
$\pt$ cut & S/B & S/$\sqrt{\rm B}$ (10 fb$^{-1}$)
    & S/$\sqrt{\rm B}$ (30 fb$^{-1}$)\\
\hline
85 GeV & 0.49(0.22)        & 6.6(4.4)       & 11.4(7.7)       \\
100 GeV & 0.61(0.26)       & 6.6(4.3)       & 11.4(7.5)        \\
\hline \hline
         &  \multicolumn{3}{c}{$m_h = 140$ GeV} \\
$\pt$ cut & S/B & S/$\sqrt{\rm B}$ (10 fb$^{-1}$)
    & S/$\sqrt{\rm B}$ (30 fb$^{-1}$)\\
\hline
85 GeV & 0.41(0.19)        & 5.5(3.7)       & 9.6(6.5)       \\
100 GeV & 0.52(0.22)       & 5.6(3.6)       & 9.6(6.3)        \\
\hline \hline
\end{tabular}
\caption{ Same with Table~\ref{table3} but at the LHC.
} \label{table4}
\end{table}

From Table~\ref{table1}-\ref{table4}, we can see that ZZ background
is greatly eliminated after including the measurement of $ZZ
\rightarrow 4\ell$. Accordingly $S/B$ and $S/\sqrt{B}$ are improved
significantly. At Tevatron, for $m_H=120$ GeV and with luminosity
$30 fb^{-1}$, over $5\sigma$ observation of Higgs boson is possible
through this channel, while one can only achieve the significance
less than $4\sigma$ without input from measurement of $ZZ
\rightarrow 4\ell$. For the heavier Higgs mass even for $m_H=140$
GeV, this channel can yield a signal of  $3\sigma$ significance with
$30 fb^{-1}$ luminosity. At LHC, our results show that for 114-140
GeV Higgs boson and with only $10 fb^{-1}$ luminosity, this channel
can provide over $5\sigma$ significance observation of invisible
Higgs boson.

\section{Discussions and open questions}

In this paper  we studied the invisible Higgs boson at Fermilab
Tevatron and CERN LHC, i.e. $ q \bar{q} \rightarrow Z H \rightarrow
\ell^+\ell^-+ \ptmiss $, where $\ptmiss$ is reconstructed from the
$\ell^+\ell^-$ with $\ell=e$ or $\mu$.
 Moreover, in
order to reduce the largest $Z(\rightarrow \ell^+\ell^-)
Z(\rightarrow \nu \bar \nu)$ background, we  propose to utilize
precise measurement of $Z(\rightarrow \ell^+\ell^-) Z(\rightarrow
\ell^+\ell^-)$ as input. Our study shows that at Tevatron, for
$m_H=120$ GeV and with luminosity $30 fb^{-1}$, over $5\sigma$
observation of Higgs boson is possible via this channel. At LHC, for
114-140 GeV Higgs boson and with only $10 fb^{-1}$ luminosity, this
channel can achieve over $5\sigma$ discovery. We should note that
the feasibility of this discovery mode depends crucially on the
understanding of the backgrounds. Therefore the simulation
incorporated with at least NLO QCD radiative correction for signal
\cite{ZHNLO} and background \cite{ZZNLO} is very important and will
put the analysis on a firmer ground. Moreover we are aware that the
simulation presented here is very rough and the full detector
simulation, which is beyond the scope of this paper, is the natural
further investigation.

It is worth to mention that invisible Higgs boson can also be
investigated through weak boson fusion (WBF) provided that events
with two energetic forward-backward jets of high dijet invariant
mass and with substantial missing transverse momentum can be
triggered efficiently \cite{Eboli:2000ze}. The study shows that it
is possible to discover invisible Higgs boson with masses up to 480
GeV at the 5 sigma level if the invisible branching ratio is close
to 100\% \cite{Eboli:2000ze}. While WBF can detect invisible Higgs
boson with higher mass, the $ZH$ channel may provide a better handle
on the mass determination \cite{Davoudiasl:2004aj}. Moreover the
combined analysis of WBF and $ZH$ modes allows a relatively
model-independent determination of invisible Higgs boson mass.

\noindent {\em Acknowledgements}: The author thanks T. Han for
correspondences on Ref.\cite{Davoudiasl:2004aj} and C. Liu for
reading the manuscript carefully. This work was supported in part by
the Natural Sciences Foundation of China under grant No. 90403004,
the trans-century fund and the key grant project (under No. 305001)
of Chinese Ministry of Education.

\end{document}